\documentclass[useAMS,usenatbib,usegraphicx]{mn2e}
\usepackage{epsf}
\usepackage{longtable}

\newcommand{\simgt}{\lower.5ex\hbox{$\; \buildrel > \over \sim \;$}}
\newcommand{\simlt}{\lower.5ex\hbox{$\; \buildrel < \over \sim \;$}}

\title[The stellar and dark matter distributions]
{The stellar and dark matter distributions in elliptical galaxies from
the ensemble of strong gravitational lenses}

\author[M.~Oguri et al.]
{Masamune Oguri,$^{1,2}$\thanks{E-mail: masamune.oguri@ipmu.jp} 
Cristian E. Rusu$^{3,4}$ and
Emilio E. Falco$^5$
\\
$^1$Department of Physics, University of Tokyo, 7-3-1 Hongo,
Bunkyo-ku, Tokyo 113-0033, Japan\\
$^2$Kavli Institute for the Physics and Mathematics of the Universe
(Kavli IPMU, WPI), University of Tokyo, Chiba 277-8583, Japan\\
$^3$Optical and Infrared Astronomy Division, National Astronomical
Observatory of Japan, 2-21-1 Osawa, Mitaka, Tokyo 181-8588, Japan\\
$^4$Department of Astronomy, University of Tokyo, 
7-3-1 Hongo, Bunkyo-ku, Tokyo 113-0033, Japan\\
$^5$Harvard-Smithsonian Center for Astrophysics, Whipple Observatory, 
670 Mt. Hopkins Road, P.O. Box 6369, Amado, AZ 85645, USA 
}

\begin{document}

\date{\today}

\voffset- .5in

\pagerange{\pageref{firstpage}--\pageref{lastpage}} \pubyear{}

\maketitle

\label{firstpage}

\begin{abstract}
We derive the average mass profile of elliptical galaxies from the
ensemble of 161 strong gravitational lens systems selected from several
surveys, assuming that the mass profile scales with the stellar mass
and effective radius of each lensing galaxy. The total mass profile is
well fitted by a power-law $\rho(r) \propto r^{\gamma}$ with 
best-fit slope $\gamma=-2.11\pm 0.05$. The decomposition of the
total mass profile into stellar and dark matter distributions is
difficult due to a fundamental degeneracy between the stellar initial
mass function (IMF) and the dark matter fraction $f_{\rm DM}$. We
demonstrate that this IMF-$f_{\rm DM}$ degeneracy can be broken by
adding direct stellar mass fraction measurements by quasar
microlensing observations. Our best-fit model prefers the Salpeter IMF
over the Chabrier IMF, and a smaller central dark matter fraction
than that predicted by adiabatic contraction models.  
\end{abstract}

\begin{keywords}
dark matter
--- galaxies: elliptical and lenticular, cD
--- galaxies: formation
--- galaxies: haloes
--- gravitational lensing: strong
\end{keywords}

\section{Introduction}

The standard collisionless cold dark matter model predicts that the
density profile of dark matter haloes is universal 
\citep[][hereafter NFW]{navarro97}. In observations, the radial
density profile of dark haloes has been tested well in clusters of
galaxies using gravitational lensing. The results indicate that the
observed radial profiles agree very well with the NFW profile from
cluster cores out to virial radii
\citep[e.g.,][]{mandelbaum06,johnston07,umetsu11,oguri12a,newman13a,okabe13}. 

On the other hand, studies of dark matter density profiles for
galaxy-scale haloes are more complicated because of the larger effects 
of central galaxies. While stacked weak lensing has shown that the 
average radial density profile of galaxy-scale haloes is consistent
with the NFW profile
\citep[e.g.,][]{hoekstra04,gavazzi07,mandelbaum08,leauthaud12}, the
inner density profile is significantly steeper than the NFW profile
due to the dominant contribution of the baryonic matter near the
galaxy centre. For example, the central density profiles of massive
elliptical galaxies have been extensively studied using velocity
dispersion measurements \citep[see][]{binney08} and strong 
gravitational lensing \citep[see][]{treu10b}, which indicates that the
total mass profiles of elliptical galaxies are nearly isothermal with
the radial density profile $\rho(r)\propto r^{-2}$.

The stellar kinematics provide a powerful means of studying the mass
profile of the core of galaxies. In particular, recent systematic
observations with integral field spectroscopy, such as SAURON
\citep{bacon01,cappellari06,cappellari07,kuntschner10} and 
ATLAS$^{\rm 3D}$
\citep{cappellari11,cappellari13a,cappellari13b,krajnovic11}, have
revealed detailed internal structures of 
elliptical galaxies. A complication in the interpretation of the
kinematics data, however, is the orbital anisotropy which is
degenerate with the mass estimate from stellar kinematics data.

Strong gravitational lensing robustly measures the projected mass
enclosed by the Einstein radius, and therefore provides a powerful
alternative to measuring the mass distribution in the cores of
elliptical galaxies. While strong lensing of single background sources
alone does not constrain the radial density profile of individual
lensing galaxies very well, the combination of strong lensing and
stellar kinematics is powerful in measuring the radial density slope,
because these two probes constrain enclosed masses at different radii
\citep{treu02,treu04}. The power of this approach has been well
demonstrated by the Sloan Lens ACS Survey 
\citep[SLACS;][]{bolton06,bolton08a,auger09,koopmans09},
the Sloan WFC Edge-on Late-type Lens Survey
 \citep[SWELLS;][]{treu11,dutton13}, and the BOSS Emission-Line Lens
 Survey \citep[BELLS;][]{brownstein12,bolton12}. Again, one of the
 major systematic uncertainties inherent in this combined analysis is
 the orbital anisotropy.

In this paper, we constrain the average mass distribution of
elliptical galaxies from the statistical analysis of a large sample of
strong gravitational lenses. Our approach is essentially similar to
the one proposed in \citet{rusin03} and \citet{rusin05} in which
a self-similar model of stellar and dark matter distributions is
assumed to describe various strong lens systems with different lens
masses and Einstein radii 
\citep[see also][]{ferreras05,bolton08b,grillo10,grillo12}.
We extend the analysis by using galaxy-galaxy strong lenses from SLACS
and BELLS as well as strongly lensed quasars, resulting in a sample
of 161 strong gravitational lenses in total. This approach relies
only on gravitational lensing, and therefore is immune to the orbital
anisotropy. 

When combining strong lenses with different masses of lensing
galaxies, we rescale the masses with stellar mass measurements of
individual lensing galaxies. However, the stellar mass estimate is
subject to various uncertainties, most notably the uncertainty
from the stellar initial mass function
\citep[IMF;][]{salpeter55,kroupa01,chabrier03} such that estimated 
stellar masses depend strongly on the assumed functional form of the
IMF. Once this uncertainty is taken into account, stellar and dark
matter distributions are highly degenerate. We break this degeneracy
using quasar microlensing \citep[see][]{wambsganss06} which directly
probes the stellar mass fraction at the positions of quasar images.

This paper is organised as follows. In Section~\ref{sec:sample} we
present our strong lens sample. We conduct a statistical analysis in
Section~\ref{sec:analysis}, and discuss implications for the adiabatic
contraction in Section~\ref{sec:ac}. Finally we summarize our results
in Section~\ref{sec:summary}. Throughout this paper we assume a flat
universe with matter density $\Omega_M=0.3$, cosmological constant 
$\Omega_\Lambda=0.7$, and Hubble constant $H_0=70\,{\rm
  km\,s^{-1}Mpc^{-1}}$.

\section{Lens Sample}\label{sec:sample}

\subsection{Strong Lenses}

Here we describe a sample of strong lenses used for our statistical
analysis. We use strong lenses discovered in various surveys. Our 
sample includes both galaxy-galaxy and quasar-galaxy lenses. In all
the samples, effective radii are measured using the de Vaucouleurs
profile.   

We use the SLACS galaxy-galaxy strong lens sample from
\citet{auger09}, in which multi-band {\it Hubble Space Telescope
  (HST)} imaging results of the full SLACS lens sample were
presented. Among the 85 grade ``A'' systems presented in
\citet{auger09}, we select a subsample of 70 lens systems based on 
the availability of multi-band images for the stellar mass estimate
(see below) and the Einstein radius measurement. Most of the lenses
were observed in the $V$- and $I$-bands, and some of them were observed in
the $B$- or $H$-bands as well. The effective radius of each lensing galaxy
measured in the $I$-band was also presented. Based on the arguments
in \citet{auger09}, we assume conservative 5\% errors on the effective
radii. The Einstein radii are derived from mass modelling assuming the
Singular Isothermal Ellipsoid (SIE) model 
\citep{bolton08a}.  

In addition, we include a galaxy-galaxy strong lens sample from BELLS
\citep{brownstein12}. We use all 25 definite lens systems, for
which the effective radius measurement from {\it HST} $I$-band
imaging data is available. We assign 10\% errors to the effective
radii \citep[see][]{brownstein12}. The Einstein radius for each lens
system was again obtained by fitting the lensed galaxy assuming the
SIE model for the mass distribution. 

We use a sample of strongly lensed quasars compiled in the
CASTLES webpage\footnote{http://cfa-www.harvard.edu/castles/}. Most of
the CASTLES quasar lenses were observed in the {\it HST} $V$-, $I$-, and
$H$-bands. We measure the effective radius (and its error) of each
lensing galaxy using the $H$-band image, or $I$-band image if the
$H$-band image is not available. We select a subsample of 38 quasar
lenses from the CASTLES based on the following criteria. 
First, both the source and lens redshifts must be measured in order to
convert the Einstein radius to the enclosed mass. Second, we exclude
complex lens systems such as lensing by multiple galaxies and lensing
by a cluster of galaxies. Third, we exclude lensing galaxies with
dominant disk components, except for Q2273+0305 which is produced by the
massive bulge of a spiral galaxy. Finally we set the condition that
the effective radius of the lensing galaxy must be measured
with a small uncertainty. We also include COSMOS5921+0638
\citep{anguita09} which is not in our CASTLES quasar lens list but has
an {\it HST} image for accurate astrometry and galaxy profile
measurements.  The resulting subsample contains 38 quasar lenses. For
each lens system we perform mass modelling using {\it glafic}
\citep{oguri10a}. We assume the SIE model plus external shear, with
priors on the ellipticity and position angle of the SIE component from
the measured galaxy light profile. We derive the Einstein radius for
each lens system from the best-fitting model.  

The Sloan Digital Sky Survey Quasar Lens Search
\citep[SQLS;][]{oguri06,oguri08,oguri12b,inada08,inada10,inada12} has
identified nearly 50 new quasar lens systems from the Sloan Digital
Sky Survey (SDSS)
data\footnote{http://www-utap.phys.s.u-tokyo.ac.jp/\~{}sdss/sqls/}. 
We are conducting a large 
programme to observe the new SQLS lens systems (\citealt{rusu11};
C. E. Rusu et al., in preparation) using the Laser Guide
Star Adaptive Optics (LGSAO) system at the Subaru telescope
\citep{hayano08,hayano10}. Among about 20 quasar lens systems we have
already observed with the Subaru LGSAO, we select a subsample of 7
quasar lenses based on the same criteria as used for the CASTLES
sample. The subsample includes SDSSJ0806+2006 which was in fact
observed with the Very Large Telescope LGSAO system \citep{sluse08}. 
Our careful analysis of the Subaru LGSAO images demonstrates that
accurate and robust estimates of galaxy morphology parameters such as
the effective radius and Sersic index are indeed feasible (C. E. Rusu
et al., in preparation). We derive the measurement error of the
effective radius for each lens system by marginalising over PSF
uncertainties. Again, the Einstein radius for each lens system is
based on mass modelling using {\it glafic} \citep{oguri10a} assuming
the SIE profile.  

The CFHTLS Strong Lensing Legacy Survey 
\citep[SL2S;][]{cabanac07,ruff11,more12,gavazzi12}
constructed a large sample of galaxy-galaxy strong lenses identified
from the Canada-France-Hawaii Telescope Legacy Survey
(CFHTLS). \citet{sonnenfeld13a,sonnenfeld13b} presented {\it HST}
imaging results and spectroscopic follow-up results for the final
sample of the SL2S galaxy-scale strong lenses. Among the 56 strong
lens candidates presented in \citet{sonnenfeld13a,sonnenfeld13b} we
select a subsample of 21 lens systems based on the availability of
{\it HST} images for determining galaxy morphology parameters as well
as spectroscopic redshifts of both the lens and source. The error on
the effective radius is conservatively assumed to be 10\%. 
\citet{sonnenfeld13a} also provided the Einstein radii derived
assuming the SIE profile. 

While the SIE profile is assumed for deriving the Einstein radius, our
conclusion is little affected by the assumption. The Einstein radius
is essentially an average distance between the lens centre and
multiple images, and is insensitive to the radial density profile,
particularly if the image configuration is symmetric. The conversion
from the Einstein radius to the two-dimensional enclosed mass within
the Einstein radius depends only on the lens and source redshifts (see
below) and is therefore model independent. For example,
\citet{jullo07} has demonstrated that the enclosed mass within the
Einstein radius is well constrained by strong lensing observations,
even if a very wide range of mass models are considered \citep[see
  also, e.g.][]{suyu12}. As a specific
example, we re-model an asymmetric lens LBQS1333+0113 using an
elliptical power-law profile with the slope $\pm0.2$ and find
that the enclosed mass is affected only by $\la 5\%$. For more
symmetric lenses this bias in mass estimates is smaller. Moreover, the
bias is essentially independent of the Einstein radius of the lens
system, and therefore its main effect is to shift the normalization
of the total mass profile, rather than biasing the radial density
slope from the combined statistical analysis. We however note that
this effect can potentially be an important source of systematic error
when analyzing larger samples of strong lenses, in case different lens
samples probing different radii are affected by this bias
differently. It is also worth noting that the mass density
profile of elliptical galaxies is not exactly a power law
\citep[e.g.,][]{chae14}.

\subsection{Stellar Mass Estimate}

We derive the stellar mass of each lensing galaxy by fitting the
observed spectral energy distribution (SED) to a stellar population
synthesis (SPS) model. Specifically, we use the SPS model of 
\citet{bruzual03}. For simplicity, we adopt a single burst model 
with the metallicity and formation redshift as parameters. We include 
a Gaussian prior on the metallicity with a mean $Z=0.01$ and standard
deviation 0.14~dex for the initial stellar mass of $10^{11}M_\odot$. 
We include the stellar mass dependence of the mean as 
$Z \propto M_*^{0.15}$. In addition, we include a Gaussian prior on
the formation redshift with a mean of 2 and standard deviation of 0.5. 
The wavelength bands for the SED fitting differ for different lens
systems, but we typically use {\it HST} $V$-, $I$-, and 
$H$-band images for the SLACS sample, {\it HST} $I$-band and SDSS
$griz$-band images for the BELLS sample, {\it HST} $V$-, $I$-, and
$H$-band images for the CASTLES sample, non-AO $I$- and AO $K'$-band
images for the SQLS+AO sample, and {\it HST} $V$-band CFHTLS
$griz$-band images for the SL2S sample. In order to accommodate the
model uncertainty, we set a minimum magnitude error of 0.1 for each
band. We assume the Salpeter IMF to derive the stellar mass, but in
our analysis below we take full account of the IMF uncertainty.

We check the validity of our stellar mass estimate by comparing our
result with stellar masses derived in \citet{auger09}. They also used
he SPS model of \citet{bruzual03}, but considered a more complex star
formation history. We find that our stellar mass estimate is in 
good agreement with that of \citet{auger09} for the same Salpeter
IMF case. More quantitatively, differences of $\log M_*^{\rm Sal}$
between our estimates and those of \citet{auger09} have a mean $-0.02$
and standard deviation $0.03$, with no clear dependence of the
difference on the stellar mass.  

\begin{figure}
\begin{center}
 \includegraphics[width=0.95\hsize]{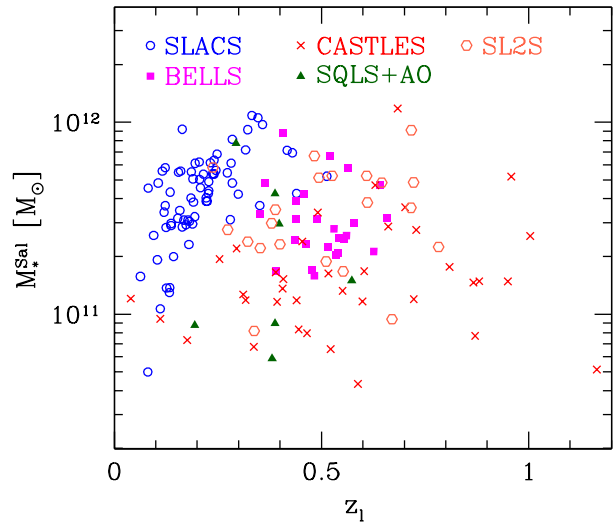}
\end{center}
\caption{Redshifts $z_l$ and stellar masses $M_*^{\rm Sal}$ derived
  assuming the Salpeter IMF for lensing galaxies of the strong lens
  sample used in our statistical analysis. Different symbols show lens
  samples from different surveys. 
\label{fig:zms}}
\end{figure}

Figure~\ref{fig:zms} shows the stellar mass and lens redshifts for the
strong lens sample. The stellar masses are in the relatively narrow
range of $10^{11}M_\odot\la M_*^{\rm Sal}\la 10^{12}M_\odot$, and the
lens redshifts are broadly distributed up to $z_l\sim 1$. 
The list of all the strong lens systems is in Appendix~\ref{sec:app}. 

\section{Statistical Analysis}\label{sec:analysis}

\subsection{Total Mass Distribution}\label{sec:plfit}

We constrain the average total mass profile for our strong lens sample
assuming that the profile scales with the stellar mass and effective
radius of the lensing galaxy. Specifically, for each strong lens
system we compute the scaled projected mass $M_{\rm tot}(<R_{\rm
  Ein})/M_*^{\rm Sal}$, where $M_{\rm tot}(<R_{\rm Ein})$ is the total
projected mass enclosed by the Einstein radius $R_{\rm
  Ein}=D_{\rm A}(z_l)\theta_{\rm Ein}$. We compute $M_{\rm
  tot}(<R_{\rm Ein})$ from the Einstein radius via $M_{\rm
  tot}(<R_{\rm Ein})=\pi R_{\rm Ein}^2\Sigma_{\rm cr}$ with
$\Sigma_{\rm cr}$ being the critical surface mass density. By
combining scaled projected mass measurements for different strong lens
systems, we can reconstruct the scaled projected mass profile, 
$M_{\rm tot}(<R)/M_*^{\rm  Sal}$, as a function of the projected
radius normalized by the effective radius, $R/R_e$
\citep{rusin03,rusin05}.  

Given the simplicity of our SPS model, errors on stellar mass
estimates from our SPS model fitting are likely to be
underestimated. We assume an error of 10\% for the stellar mass
estimate for the  SLACS, CASTLES, and SQLS+AO samples, and a larger
error of 20\% for the BELLS and SL2S samples given the lack of
high-resolution near-infrared images. The assumed model error of the
stellar masses for the SLACS sample is comparable to the estimate
in \citet{auger09} in which more complex SPS model was considered. The
measurement error of the effective radius $R_e$ is propagated to the
error on the total projected mass assuming $M_{\rm tot}(<R) \propto
R$, which will be shown to be reasonable in our analysis below. We
neglect the error on the Einstein radius because it is usually much
smaller compared with errors on the stellar mass and effective radius.

\begin{figure}
\begin{center}
 \includegraphics[width=0.95\hsize]{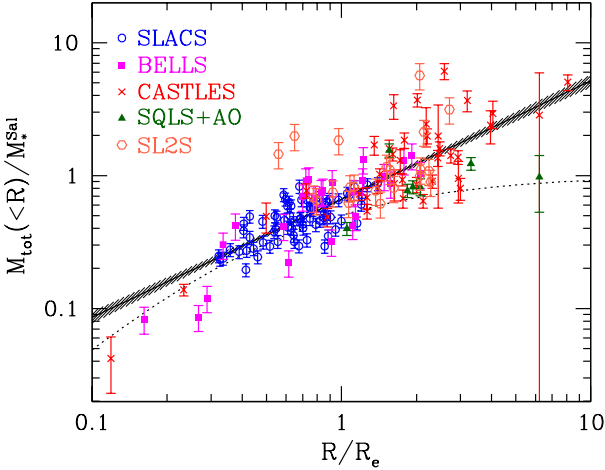}
\end{center}
\caption{The scaled projected mass, $M_{\rm tot}(<R)/M_*^{\rm Sal}$, as
  a function of the projected radius normalized by the effective
  radius, $R/R_e$ for our sample of strong lenses. The solid line
  shows the power-law fit (equation~\ref{eq:powerlaw}), and the shaded
  region indicates the 1$\sigma$ range. For reference, we show the
  projected stellar mass profile for the Salpeter IMF as the dotted
  line.  
\label{fig:mtot}}
\end{figure}

Figure~\ref{fig:mtot} shows projected mass measurements for our sample
of strong lenses. There is a clear trend that the scaled projected
mass is roughly proportional to the projected radius. We fit the trend
with a self-similar power-law mass model of the following form:
\begin{equation}
\frac{M_{\rm tot}(<R)}{M_*^{\rm
    Sal}}=A\left(\frac{R}{R_e}\right)^{3+\gamma},
\label{eq:powerlaw}
\end{equation}
where $\gamma$ is the radial slope of the three-dimensional density
profile, $\rho(r)\propto r^{\gamma}$. The relation above implicitly
assumes spherical symmetry for the lens. The elongation of lens galaxy
shapes along the line-of-sight induces an additional error, which
averages out when we combine many strong lens systems. Following
\citet{rusin03}, in what follows we take account of the diversity of
individual lens systems, such as the non-sphericity and scatters in
radial slopes and dark matter fractions, by uniformly rescaling the
estimated errors by a constant factor ($\sigma\rightarrow f\sigma$
with $f\sim 2.9$ in our analysis) so that the best-fit model has
$\chi^2/N_{\rm dof}=1$, where $N_{\rm dof}$ is the number of degrees
of freedom. We find the best-fit slope of $\gamma=-2.11\pm0.05$, which
is slightly steeper than the singular isothermal profile
($\gamma=-2$). Our result is consistent with earlier attempts by
\citet{rusin03} and \citet{rusin05}, and also with the combined
lensing and kinematics constraints \citep{koopmans09,auger10b,bolton12}.  

\subsection{The IMF-$f_{\rm DM}$ degeneracy}
\label{sec:imf-fdm}

We now want to decompose the total mass distribution into the stellar
and dark matter distributions.  
However there is a fundamental difficulty in this decomposition,
because of the well-known degeneracy between the relative
contributions of stellar and dark matter. This can be understood very
easily; if we ignore a minimum stellar $M/L$, all the observed lensing
and kinematics data should be explained by dark matter only, without
adding any contributions from stellar masses, as long as the assumed
dark matter distribution is flexible enough. This indicates that IMF
models that predict very small stellar masses cannot be excluded if we
allow enough dark matter in galaxy cores to explain lensing and
kinematics data, suggesting a fundamental degeneracy between the IMF
and the dark matter fraction $f_{\rm DM}$. Hereafter we refer to it as
the IMF-$f_{\rm DM}$ degeneracy.   

How can we break the IMF-$f_{\rm DM}$ degeneracy? The traditional
approach is to add priors on the dark matter distribution
\citep[e.g.,][]{treu10a,auger10a,dutton11,cappellari12,sonnenfeld12}
and the population of dark haloes \citep{dutton13}. For instance, the
dark matter distribution is often assumed to follow the NFW 
profile or the NFW profile modified by the so-called adiabatic
contraction \citep{blumenthal86,gnedin04,abadi10}. Sometimes it is
also assumed that the contribution of dark matter is negligibly small
at the core of galaxies. 

Alternatively, the degeneracy is broken if we add an independent
constraint on the IMF. Indeed such a constraint is available from 
the spectral features that are sensitive to dwarf stars 
\citep{vandokkum10,conroy12,ferreras13,conroy13,spiniello13}.
These analyses indicate that the IMF of massive elliptical galaxies
tends to have a Salpeter-like ``bottom-heavy'' shape and disfavors
Chabrier-like IMFs. Recently \citet{barnabe13} combined such
spectroscopic analysis with gravitational lensing and dynamical data
to break the IMF-$f_{\rm DM}$ degeneracy and to constrain the shape of
the IMF.

In this paper, we employ a totally different approach to break the
IMF-$f_{\rm DM}$ degeneracy. The idea is to add constraints from
quasar microlensing which directly measures the fraction of mass in
the form of stars (stellar mass fraction) at the positions of lensed
quasar images 
\citep[e.g.,][]{schechter02,kochanek04,bate07,pooley09,mediavilla09}. 
Simply stated, we can constrain the stellar mass fraction because the
effect of microlensing is more pronounced for a higher stellar mass
fraction. Thus, quasar microlensing measurements directly constrain
$f_{\rm DM}$ at the projected positions of lensed quasar images and
hence break the IMF-$f_{\rm DM}$ degeneracy.  

Specifically we adopt X-ray microlensing measurements of 12 quasar
lenses presented in \citet{pooley12} as well as optical microlensing
measurements of 3 quasar lenses presented in \citet{bate11}. We
exclude H1413+117 in the X-ray microlensing sample of
\citet{pooley12} from our analysis because the effective radius is 
not measured for this lens system. In our analysis we adopt the
probability distributions of the stellar mass fraction obtained for
individual lens systems and include them as constraints at radii
$R_{\rm Ein}/R_e$. Here we ignore the effect of different $R/R_e$ for
different quasar images for a given strong lens system. X-ray quasar
microlensing has an advantage over optical microlensing in that the
size of the X-ray emitting region is much smaller than the Einstein
radii of stars  
\citep{pooley07,morgan08,morgan12,chartas09,chartas12,dai10}  and
hence results are insensitive to the assumed source sizes. On the
other hand, the optical microlensing results of \citet{bate11} involve
a proper marginalization over the size of the optical emitting region. 
We note that the microlensing measurements of the stellar mass
fraction are not very sensitive to the slope of the IMF
\citep{wyithe01,schechter04,congdon07}, and therefore we can assume
that the microlensing constraints on the stellar mass fraction are
independent of the IMF, at least for our range of interest.

\subsection{Two Components Model}

Next we consider a two-component model that consists of stellar and
dark matter. The stellar matter component is modelled by the Hernquist
profile \citep{hernquist90} that resembles the de Vaucouleurs profile
when projected along the line-of-sight. Specifically, we model the
enclosed projected mass profile of the stellar component as 
\citep{keeton01}
\begin{equation}
\frac{M_{\rm ste}(<R)}{M_*^{\rm Sal}}=\alpha_{\rm SPS}^{\rm Sal}
\left(\frac{R}{R_b}\right)^2\frac{1-F(R/R_b)}{(R/R_b)^2-1},
\end{equation}
\begin{eqnarray}
 F(u)&=&\frac{1}{\sqrt{1-u^2}}{\rm arctanh}\sqrt{1-u^2}\;\;\;\;(u<1),\\ 
&=& \frac{1}{\sqrt{u^2-1}}{\rm arctan}\sqrt{u^2-1}\;\;\;\;(u>1).
\end{eqnarray}
with $R_b=0.551R_e$. The parameter $\alpha_{\rm SPS}^{\rm Sal}$ takes
account of the uncertainty of the IMF; $\alpha_{\rm SPS}^{\rm Sal}=1$
means that the IMF is described by the Salpeter IMF, whereas 
$\alpha_{\rm SPS}^{\rm Sal}\approx 0.56$ corresponds to the Chabrier
IMF. On the other hand, we assume a simple power-law mass distribution
for the dark matter distribution
\begin{equation}
\frac{M_{\rm DM}(<R)}{M_*^{\rm Sal}}=A_{\rm
  DM}\left(\frac{R}{R_e}\right)^{3+\gamma_{\rm DM}}.
\end{equation}
The total mass distribution is simply given by the sum of these two
components 
\begin{equation}
\frac{M_{\rm tot}(<R)}{M_*^{\rm Sal}}=
\frac{M_{\rm ste}(<R)}{M_*^{\rm Sal}}+\frac{M_{\rm DM}(<R)}{M_*^{\rm Sal}}.
\end{equation}
Thus the two-component model has 3 parameters, $\alpha_{\rm SPS}^{\rm
 Sal}$, $A_{\rm DM}$, and $\gamma_{\rm DM}$.

\begin{figure}
\begin{center}
 \includegraphics[width=0.95\hsize]{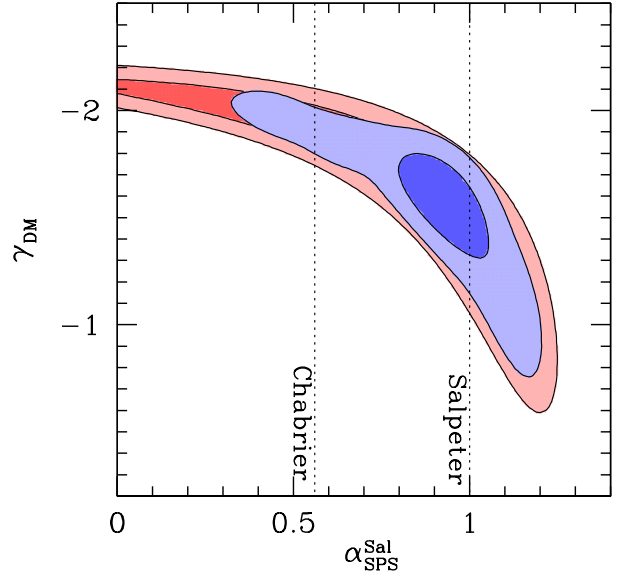}
\end{center}
\caption{Projected constraints in the $\alpha_{\rm SPS}^{\rm
 Sal}$-$\gamma_{\rm DM}$ plane for the two-component model. Outer
  ({\it red}) contours show 1 and 2$\sigma$ contours without adding
  the microlensing constraints (see Section~\ref{sec:imf-fdm}),
  whereas the inner ({\it blue}) contours show the constraints
  including the microlensing constraints. Values of  $\alpha_{\rm
    SPS}^{\rm Sal}$ corresponding to the Salpeter and Chabrier IMFs
  are indicated by vertical dotted lines. 
\label{fig:cont_tc_alpgam}}
\end{figure}

\begin{figure}
\begin{center}
 \includegraphics[width=0.95\hsize]{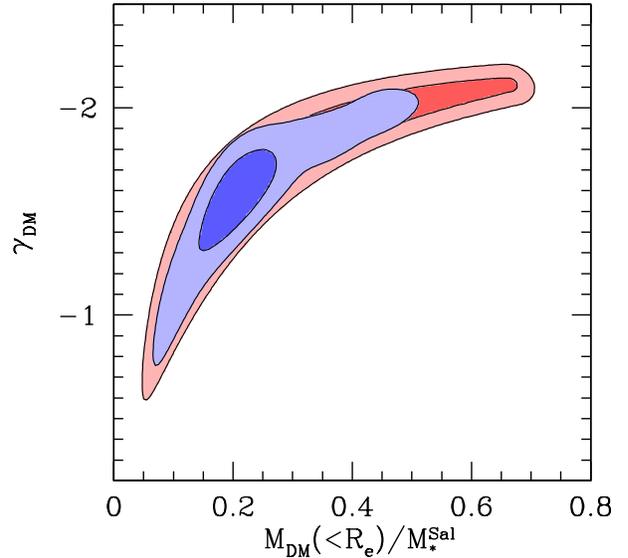}
\end{center}
\caption{Similar to Figure~\ref{fig:cont_tc_alpgam}, but the projected
  constraints in the $A_{\rm DM}=M_{\rm DM}(<R_e)/M_*^{\rm
    Sal}$-$\gamma_{\rm DM}$ plane are shown.
\label{fig:cont_tc_gamnorm}}
\end{figure}

\begin{figure}
\begin{center}
 \includegraphics[width=0.95\hsize]{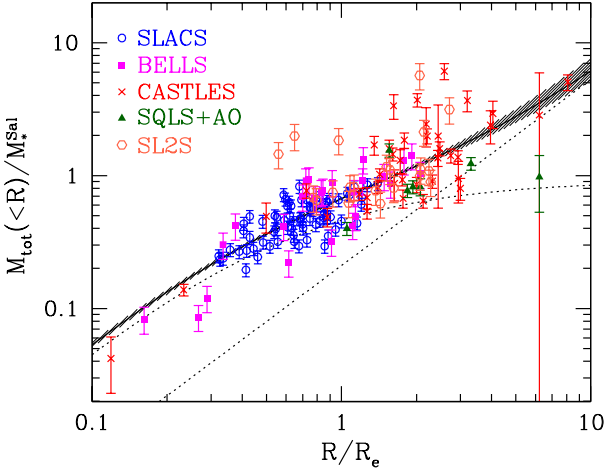}
\end{center}
\caption{Similar to Figure~\ref{fig:mtot}, but the best-fit
  two-component model is overplotted. The solid line with shading
  shows the best-fit and $1\sigma$ range of the total mass
  profile. Dotted lines indicate best-fit stellar and dark matter
  distributions. 
\label{fig:mtot2}}
\end{figure}

Figures~\ref{fig:cont_tc_alpgam} and \ref{fig:cont_tc_gamnorm} show
projected constraints in the $\alpha_{\rm SPS}^{\rm Sal}$-$\gamma_{\rm DM}$ 
and $A_{\rm DM}$-$\gamma_{\rm DM}$ planes, respectively. Without the
microlensing constraints (see Section~\ref{sec:imf-fdm} for details),
the constraints are quite degenerate such that models with
$\alpha_{\rm SPS}^{\rm Sal}\approx 0$ are allowed. Our result
indicates that the quasar microlensing measurements of the stellar
mass fraction indeed break the IMF-$f_{\rm DM}$ degeneracy. The
best-fit parameters are $\alpha_{\rm SPS}^{\rm
  Sal}=0.92_{-0.08}^{+0.09}$, 
$\gamma_{\rm DM}=-1.60_{-0.13}^{+0.18}$, 
and $A_{\rm DM}=M_{\rm DM}(<R_e)/M_*^{\rm Sal}=0.21\pm0.04$. The
Salpeter IMF is preferred over the Chabrier IMF, which is in line with
recent claims based on subtle spectral features \citep[][but see also
  \citealt{ferreras08,ferreras10,smith13}]{vandokkum10,conroy12,ferreras13,conroy13,spiniello13}. 
In addition, we find that models without dark matter ($A_{\rm DM}=0$)
are disfavored at the $5\sigma$ level even without the microlensing
constraints. The best-fit two component model is shown in
Figure~\ref{fig:mtot2}. 

\subsection{Mass and Redshift Dependences}\label{sec:evo}

There have been several indications from recent lensing and/or
kinematics studies \citep{treu10a,dutton11,dutton12,cappellari12} as
well as from studies of spectral features
\citep{vandokkum10,conroy12,ferreras13,conroy13,spiniello13} that the
IMF is non-universal, i.e., the IMF changes with galaxy velocity
dispersions and stellar masses. Some previous studies from combined
lensing kinematics analyses have also indicated possible redshift
evolution of the slope of the total mass profile
\citep{ruff11,bolton12,sonnenfeld13b}.

Here we investigate whether the total mass profile measured from the
ensemble of strong lenses depends on the stellar mass or the redshift.
We divide our strong lens sample into subsamples of different stellar
mass or redshift bins to see how the fitting parameters change with
these parameters. Specifically, we consider two stellar mass bins
divided at $M_*^{\rm Sal}=3\times 10^{11}M_\odot$ and two redshift
bins divided at $z_l=0.4$. For each subsample we repeat the power-law
fit to the total mass profile as presented in Section~\ref{sec:plfit},
and derive constraints on the mass normalization $A$ and the radial
slope $\gamma$ in equation~(\ref{eq:powerlaw}).

\begin{figure}
\begin{center}
 \includegraphics[width=0.95\hsize]{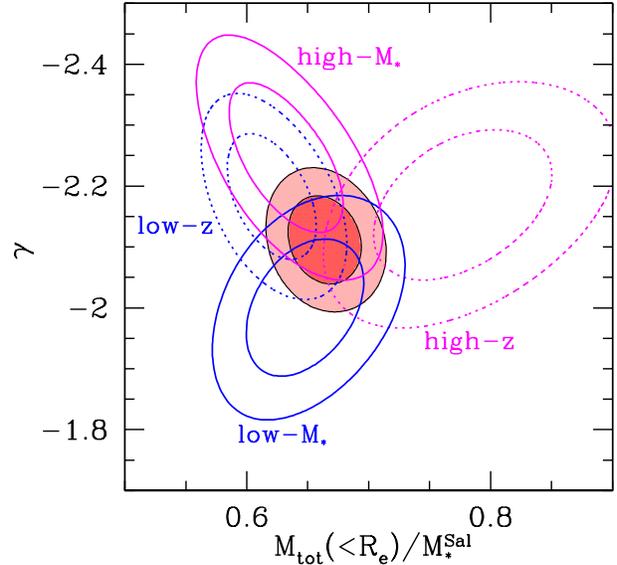}
\end{center}
\caption{Constraints in the $A=M_{\rm tot}(<R_e)/M_*^{\rm
    Sal}$-$\gamma$ plane for the power-law model (see
  Section~\ref{sec:plfit}). Filled contours show 1 and 2$\sigma$
  contours from the full strong lens sample. Contours with solid
  lines show 1 and 2$\sigma$ contours from subsamples with stellar
  mass $M_*^{\rm Sal}$ larger or smaller than $3\times
  10^{11}M_\odot$. Contours with dotted lines show 1 and 2$\sigma$
  contours from subsamples with redshift lower or higher than
  $0.4$. 
\label{fig:cont_pl}}
\end{figure}

Figure~\ref{fig:cont_pl} shows constraints in the $A=M_{\rm
  tot}(<R_e)/M_*^{\rm Sal}$-$\gamma$ plane. We find trends of the
best-fit values, such that the higher stellar mass sample prefers
steeper radial slope, and the higher redshift sample prefers larger
normalization of the total mass profile. One possible interpretation
of the dependence on the stellar mass is that the lower stellar mass
sample has a larger satellite fraction and therefore effectively
shallower radial density slope. The larger mass normalization for the
higher redshift sample can be due to either a larger dark matter
fraction or a larger stellar mass (i.e., larger $\alpha_{\rm
  SPS}^{\rm Sal}$). The larger dark matter fraction at higher redshift
may be explained by star formation in these galaxies or infall of
satellite galaxies via dynamical friction. 
We note however that these trends with the stellar mass and redshift
are not very significant, at $\la 2\sigma$ level. Improved statistical
analysis with a significantly larger sample of strong gravitational
lenses is necessary for more detailed studies.

\section{Implications for the Adiabatic Contraction}\label{sec:ac}

Our measurements of the average dark matter distribution at the core
of elliptical galaxies enable a direct test of models of the
modification of the dark matter density profile due to baryonic
physics. The most popular model of such a baryonic effect has been the
adiabatic contraction \citep{blumenthal86,gnedin04,abadi10} which
predicts that the dissipative collapse of baryons leads to a more 
centrally concentrated dark matter distribution as compared with what
we would expect for the case of no baryons.

Here we compute the expected dark matter distribution for our sample of
strong lenses, as follows. We employ the stellar mass-dark halo
relation derived in \citet{leauthaud12} in which the relation has
been constrained up to $z\sim 1$ from lensing and clustering
observations. We note that \citet{leauthaud12} assumed the Chabrier
IMF for computing the stellar mass, and thus $M_*^{\rm Sal}$ for our
lens sample is first converted to stellar mass with the Chabrier
IMF by multiplying by 0.56 before applying the stellar mass-dark halo
relation in \citet{leauthaud12} to compute the halo mass for each lens 
system. We adopt the mass-concentration relation of \citet{duffy08} to
compute the concentration parameter. We add a log-normal scatter of
0.2~dex for the concentration parameter, the baryon mass fraction, and 
the scale radius. Again we assume the Hernquist model for the baryon
mass distribution. We assume that the dark matter distribution is
modified from the NFW profile by the adiabatic contraction model of
\citet{gnedin04} who derived a fitting formula of the adiabatic
contraction based on high-resolution hydrodynamical simulations. 
For comparison, we consider a dark matter model of the pure NFW
profile without the adiabatic contraction. 

\begin{figure}
\begin{center}
 \includegraphics[width=0.95\hsize]{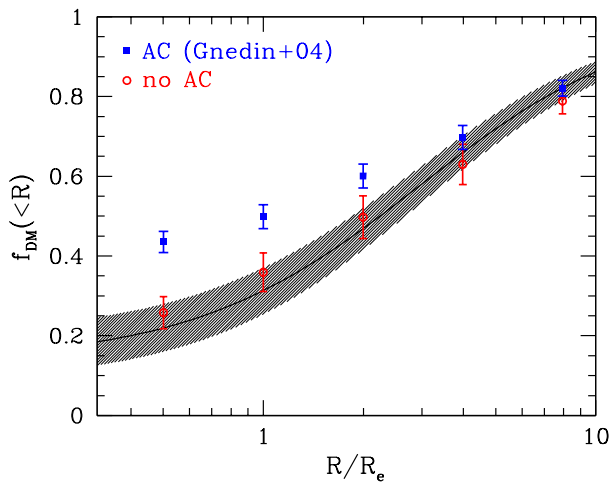}
\end{center}
\caption{The dark matter fraction within the projected radius $R$. The
  solid line with shading shows the dark matter fraction from the
  best-fit two component model. Filled squares with errors indicate the
  expected dark matter fraction for the adiabatic contraction model of 
  \citet{gnedin04}. Open circles with errors show the expected dark
  matter fraction for the NFW profile without adiabatic contraction. 
\label{fig:fdm}}
\end{figure}

Figure~\ref{fig:fdm} compares the dark matter fractions at several
different radii for the best-fit two component model with the model
predictions described above. We find that the adiabatic contraction
model of \citet{gnedin04} overpredicts the dark matter fraction at
$R\la 2R_e$. The observed dark matter fraction is more consistent with
the NFW model without the adiabatic contraction. 

Our result is in line with recent studies with lensing and stellar
kinematics which prefer moderate or no adiabatic contraction
\citep[e.g.,][]{auger10a,dutton13,newman13b}, and suggests other
physical processes may also play an important role. For instance,
dissipationless mergers of stellar clumps can indeed decrease the
central dark matter density as compared with the one predicted by 
adiabatic contraction
\citep[e.g.,][]{naab07,lackner10}. Dissipationless mergers appear to
be consistent with the possible redshift evolution of the dark matter
fraction as discussed in Section~\ref{sec:evo}. The effect of baryon
mass loss induced by feedbacks can also counteract the adiabatic
contraction \citep{ragone12}. 

\section{Summary}\label{sec:summary}

We have studied the average mass distribution of elliptical galaxies
with the statistical analysis of 161 strong gravitational lens
systems compiled from several surveys. Each strong lens system
provides a robust measurement of the enclosed mass within the Einstein
radius, and hence assuming that the mass distribution scales with the
stellar mass and the effective radius we can reconstruct the total
mass distribution. When fitted to a single power-law, the total mass
profile is described by $\rho(r) \propto r^{\gamma}$ with the
best-fit slope of $\gamma=-2.11\pm 0.05$.  We have argued that the
decomposition of the total mass profile into the stellar and dark
matter distribution involves a fundamental difficulty due to the
IMF-$f_{\rm DM}$ degeneracy, which is very difficult to break if we
assume flexible enough dark matter distributions. We have demonstrated
that the IMF-$f_{\rm DM}$ degeneracy can be broken by adding quasar
microlensing constraints which directly measure the stellar mass
fraction at the positions of lensed quasar images. Our best-fit model
favors the Salpeter IMF over the Chabrier IMF and the best-fit dark
matter density slope of $\gamma_{\rm DM}=-1.60_{-0.13}^{+0.18}$. The
inclusion of dark matter component is required at the $5\sigma$ level. 
We identify possible trends of the total density profile with the
stellar mass and redshift. Finally, we have compared the observed dark
matter fraction with predicted dark matter fractions with and without
adiabatic contraction and found that the model without adiabatic
contraction better explains the result. 

These results are obtained using gravitational lensing only without
relying on the stellar kinematics data. Our results are generally in
agreement with results using the stellar kinematics information, in
which nearly isotropic velocity dispersion near the galaxy center is
assumed. This suggests that the velocity anisotropy is indeed small,
although careful combined analysis will be necessary to assess the
degree of the velocity anisotropy more quantitatively.

A more comprehensive analysis of dependences of the stellar and dark
matter distributions will require a larger sample of galaxy scale
strong gravitational lenses, which will be obtained in future
wide-field imaging surveys \citep[e.g.,][]{oguri10b}. Large
samples of strong lenses are being constructed from bright
submillimeter galaxies
\citep{negrello10,gonzalez12,vieira13,bussmann13}, which should 
significantly advance various statistical analyses of strong
gravitational lenses as the one presented in this paper. Measurements
of quasar microlensing for more quasar lens systems are also important
to map the dark matter content of galaxies accurately and to reduce
potential systematic errors associated with the use of a subsample of
strong lens systems for quasar microlensing constraints.

\section*{Acknowledgments}
We thank Masataka Fukugita and Surhud More for useful discussions, and
Dominique Sluse for comments. We also thank an anonymous referee for
useful suggestions. This work was supported in part by the FIRST program
``Subaru Measurements of Images and Redshifts (SuMIRe)'', World
Premier International Research Center Initiative (WPI Initiative),
MEXT, Japan, and Grant-in-Aid for Scientific Research from the JSPS 
(23740161). 


\appendix

\section{List of strong gravitational lenses}\label{sec:app}

Table~\ref{tab:app} shows a list of all 161 strong gravitational lens
systems used in this paper.

\onecolumn
\setlongtables
\begin{longtable}{ccccccc}
 \caption{List of 161 strong gravitational lens systems used for the
   statistical analysis. For each strong lens system we show the lens
   redshift $z_l$, the source redshift ($z_s$), the effective radius
   measured at the intermediate axis ($\theta_e$), the stellar mass
   assuming the Salpeter IMF ($\log M_*^{\rm Sal}$), and the Einstein
   radius ($\theta_{\rm Ein}$). 
\label{tab:app}}\\
  \hline
   Name  & Sample & $z_l$ & $z_s$ & $\theta_e$ & $\log M_*^{\rm Sal}$ & $\theta_{\rm Ein}$\\
   & & & & (arcsec) & ($M_\odot$) & (arcsec) \\
 \hline
 \endfirsthead
\multicolumn{7}{c}{Table~\ref{tab:app} continued}\\
\hline\hline
   Name  & Sample & $z_l$ & $z_s$ & $\theta_e$ & $\log M_*^{\rm Sal}$ & $\theta_{\rm Ein}$\\
   & & & & (arcsec) & ($M_\odot$) & (arcsec) \\
\hline
\endhead
\hline
\endfoot
\hline
\endlastfoot
SDSSJ0008$-$0004 &   SLACS & 0.440 & 1.192 & $1.71\pm0.09$ & 11.63 & 1.16 \\
SDSSJ0029$-$0055 &   SLACS & 0.227 & 0.931 & $2.16\pm0.11$ & 11.63 & 0.96 \\
SDSSJ0037$-$0942 &   SLACS & 0.195 & 0.632 & $1.80\pm0.09$ & 11.79 & 1.53 \\
  SDSSJ0044+0113 &   SLACS & 0.120 & 0.197 & $1.92\pm0.10$ & 11.53 & 0.80 \\
SDSSJ0157$-$0056 &   SLACS & 0.513 & 0.924 & $1.84\pm0.09$ & 11.72 & 0.79 \\
SDSSJ0216$-$0813 &   SLACS & 0.332 & 0.523 & $2.40\pm0.12$ & 12.04 & 1.16 \\
  SDSSJ0252+0039 &   SLACS & 0.280 & 0.982 & $1.39\pm0.07$ & 11.49 & 1.04 \\
SDSSJ0330$-$0020 &   SLACS & 0.351 & 1.071 & $0.91\pm0.05$ & 11.57 & 1.10 \\
  SDSSJ0728+3835 &   SLACS & 0.206 & 0.688 & $1.78\pm0.09$ & 11.70 & 1.25 \\
  SDSSJ0737+3216 &   SLACS & 0.322 & 0.581 & $1.80\pm0.09$ & 11.96 & 1.00 \\
  SDSSJ0819+4534 &   SLACS & 0.194 & 0.446 & $1.98\pm0.10$ & 11.51 & 0.85 \\
  SDSSJ0822+2652 &   SLACS & 0.241 & 0.594 & $1.82\pm0.09$ & 11.73 & 1.17 \\
  SDSSJ0841+3824 &   SLACS & 0.116 & 0.657 & $4.21\pm0.21$ & 11.75 & 1.41 \\
  SDSSJ0903+4116 &   SLACS & 0.430 & 1.065 & $1.78\pm0.09$ & 11.84 & 1.29 \\
  SDSSJ0912+0029 &   SLACS & 0.164 & 0.324 & $4.01\pm0.20$ & 11.96 & 1.63 \\
SDSSJ0935$-$0003 &   SLACS & 0.347 & 0.467 & $2.15\pm0.11$ & 12.02 & 0.87 \\
  SDSSJ0936+0913 &   SLACS & 0.190 & 0.588 & $2.11\pm0.11$ & 11.71 & 1.09 \\
  SDSSJ0946+1006 &   SLACS & 0.222 & 0.609 & $2.35\pm0.12$ & 11.61 & 1.38 \\
  SDSSJ0955+0101 &   SLACS & 0.111 & 0.316 & $1.47\pm0.07$ & 11.03 & 0.91 \\
  SDSSJ0956+5100 &   SLACS & 0.241 & 0.470 & $2.19\pm0.11$ & 11.80 & 1.33 \\
  SDSSJ0959+4416 &   SLACS & 0.237 & 0.531 & $1.98\pm0.10$ & 11.73 & 0.96 \\
  SDSSJ0959+0410 &   SLACS & 0.126 & 0.535 & $1.29\pm0.06$ & 11.14 & 0.99 \\
  SDSSJ1016+3859 &   SLACS & 0.168 & 0.439 & $1.46\pm0.07$ & 11.49 & 1.09 \\
  SDSSJ1020+1122 &   SLACS & 0.282 & 0.553 & $1.59\pm0.08$ & 11.79 & 1.20 \\
  SDSSJ1023+4230 &   SLACS & 0.191 & 0.696 & $1.77\pm0.09$ & 11.58 & 1.41 \\
  SDSSJ1029+0420 &   SLACS & 0.104 & 0.615 & $1.56\pm0.08$ & 11.28 & 1.01 \\
  SDSSJ1032+5322 &   SLACS & 0.133 & 0.329 & $0.81\pm0.04$ & 11.11 & 1.03 \\
  SDSSJ1100+5329 &   SLACS & 0.317 & 0.858 & $2.20\pm0.11$ & 11.86 & 1.52 \\
  SDSSJ1103+5322 &   SLACS & 0.158 & 0.735 & $2.85\pm0.14$ & 11.54 & 1.02 \\
  SDSSJ1106+5228 &   SLACS & 0.095 & 0.407 & $1.39\pm0.07$ & 11.41 & 1.23 \\
  SDSSJ1112+0826 &   SLACS & 0.273 & 0.629 & $1.32\pm0.07$ & 11.74 & 1.49 \\
  SDSSJ1134+6027 &   SLACS & 0.153 & 0.474 & $2.02\pm0.10$ & 11.50 & 1.10 \\
  SDSSJ1142+1001 &   SLACS & 0.222 & 0.504 & $1.24\pm0.06$ & 11.59 & 0.98 \\
SDSSJ1143$-$0144 &   SLACS & 0.106 & 0.402 & $2.66\pm0.13$ & 11.68 & 1.68 \\
  SDSSJ1153+4612 &   SLACS & 0.180 & 0.875 & $1.16\pm0.06$ & 11.36 & 1.05 \\
  SDSSJ1204+0358 &   SLACS & 0.164 & 0.631 & $1.09\pm0.05$ & 11.45 & 1.31 \\
  SDSSJ1205+4910 &   SLACS & 0.215 & 0.481 & $1.79\pm0.09$ & 11.73 & 1.22 \\
  SDSSJ1213+6708 &   SLACS & 0.123 & 0.640 & $1.50\pm0.08$ & 11.57 & 1.42 \\
  SDSSJ1218+0830 &   SLACS & 0.135 & 0.717 & $2.70\pm0.14$ & 11.64 & 1.45 \\
  SDSSJ1250+0523 &   SLACS & 0.232 & 0.795 & $1.32\pm0.07$ & 11.79 & 1.13 \\
SDSSJ1251$-$0208 &   SLACS & 0.224 & 0.784 & $2.61\pm0.13$ & 11.59 & 0.84 \\
  SDSSJ1306+0600 &   SLACS & 0.173 & 0.472 & $1.25\pm0.06$ & 11.47 & 1.32 \\
  SDSSJ1313+4615 &   SLACS & 0.185 & 0.514 & $1.59\pm0.08$ & 11.61 & 1.37 \\
SDSSJ1318$-$0313 &   SLACS & 0.240 & 1.300 & $2.51\pm0.13$ & 11.73 & 1.58 \\
SDSSJ1330$-$0148 &   SLACS & 0.081 & 0.711 & $0.96\pm0.05$ & 10.70 & 0.86 \\
  SDSSJ1402+6321 &   SLACS & 0.205 & 0.481 & $2.29\pm0.11$ & 11.79 & 1.35 \\
  SDSSJ1403+0006 &   SLACS & 0.189 & 0.473 & $1.14\pm0.06$ & 11.47 & 0.83 \\
  SDSSJ1416+5136 &   SLACS & 0.299 & 0.811 & $0.98\pm0.05$ & 11.63 & 1.37 \\
  SDSSJ1420+6019 &   SLACS & 0.063 & 0.535 & $2.25\pm0.11$ & 11.20 & 1.04 \\
  SDSSJ1430+4105 &   SLACS & 0.285 & 0.575 & $2.55\pm0.13$ & 11.91 & 1.52 \\
  SDSSJ1432+6317 &   SLACS & 0.123 & 0.664 & $3.04\pm0.15$ & 11.76 & 1.26 \\
SDSSJ1436$-$0000 &   SLACS & 0.285 & 0.805 & $1.63\pm0.08$ & 11.68 & 1.12 \\
  SDSSJ1443+0304 &   SLACS & 0.134 & 0.419 & $0.70\pm0.03$ & 11.14 & 0.81 \\
SDSSJ1451$-$0239 &   SLACS & 0.125 & 0.520 & $1.54\pm0.08$ & 11.45 & 1.04 \\
  SDSSJ1525+3327 &   SLACS & 0.358 & 0.717 & $2.42\pm0.12$ & 11.99 & 1.31 \\
SDSSJ1531$-$0105 &   SLACS & 0.160 & 0.744 & $1.97\pm0.10$ & 11.75 & 1.71 \\
  SDSSJ1538+5817 &   SLACS & 0.143 & 0.531 & $1.00\pm0.05$ & 11.30 & 1.00 \\
  SDSSJ1614+4522 &   SLACS & 0.178 & 0.811 & $2.58\pm0.13$ & 11.49 & 0.84 \\
  SDSSJ1621+3931 &   SLACS & 0.245 & 0.602 & $1.51\pm0.08$ & 11.75 & 1.29 \\
SDSSJ1627$-$0053 &   SLACS & 0.208 & 0.524 & $1.98\pm0.10$ & 11.66 & 1.23 \\
  SDSSJ1630+4520 &   SLACS & 0.248 & 0.793 & $1.65\pm0.08$ & 11.83 & 1.78 \\
  SDSSJ1636+4707 &   SLACS & 0.228 & 0.675 & $1.68\pm0.08$ & 11.69 & 1.08 \\
  SDSSJ1644+2625 &   SLACS & 0.137 & 0.610 & $1.55\pm0.08$ & 11.46 & 1.27 \\
  SDSSJ1719+2939 &   SLACS & 0.181 & 0.578 & $1.46\pm0.07$ & 11.48 & 1.28 \\
SDSSJ2238$-$0754 &   SLACS & 0.137 & 0.713 & $1.82\pm0.09$ & 11.47 & 1.27 \\
  SDSSJ2300+0022 &   SLACS & 0.228 & 0.463 & $1.52\pm0.08$ & 11.64 & 1.24 \\
  SDSSJ2303+1422 &   SLACS & 0.155 & 0.517 & $2.94\pm0.15$ & 11.74 & 1.62 \\
SDSSJ2321$-$0939 &   SLACS & 0.082 & 0.532 & $4.11\pm0.21$ & 11.66 & 1.60 \\
  SDSSJ2341+0000 &   SLACS & 0.186 & 0.807 & $2.36\pm0.12$ & 11.74 & 1.44 \\
SDSSJ2347$-$0005 &   SLACS & 0.417 & 0.714 & $1.14\pm0.06$ & 11.85 & 1.11 \\
  SDSSJ0151+0049 &   BELLS & 0.517 & 1.364 & $0.67\pm0.07$ & 11.35 & 0.75 \\
  SDSSJ0747+5055 &   BELLS & 0.438 & 0.898 & $1.09\pm0.11$ & 11.50 & 0.64 \\
  SDSSJ0747+4448 &   BELLS & 0.437 & 0.897 & $0.92\pm0.09$ & 11.39 & 0.72 \\
  SDSSJ0801+4727 &   BELLS & 0.483 & 1.518 & $0.50\pm0.05$ & 11.20 & 0.89 \\
  SDSSJ0830+5116 &   BELLS & 0.530 & 1.332 & $0.97\pm0.10$ & 11.44 & 0.89 \\
SDSSJ0944$-$0147 &   BELLS & 0.539 & 1.179 & $0.48\pm0.05$ & 11.32 & 0.92 \\
SDSSJ1159$-$0007 &   BELLS & 0.579 & 1.346 & $0.96\pm0.10$ & 11.48 & 0.81 \\
  SDSSJ1215+0047 &   BELLS & 0.642 & 1.297 & $0.65\pm0.07$ & 11.67 & 0.74 \\
  SDSSJ1221+3806 &   BELLS & 0.535 & 1.284 & $0.47\pm0.05$ & 11.31 & 0.74 \\
SDSSJ1234$-$0241 &   BELLS & 0.490 & 1.016 & $1.05\pm0.11$ & 11.50 & 0.28 \\
SDSSJ1318$-$0104 &   BELLS & 0.659 & 1.396 & $0.69\pm0.07$ & 11.50 & 0.84 \\
  SDSSJ1337+3620 &   BELLS & 0.564 & 1.182 & $2.03\pm0.20$ & 11.76 & 0.68 \\
  SDSSJ1349+3612 &   BELLS & 0.440 & 0.893 & $1.89\pm0.19$ & 11.59 & 0.71 \\
  SDSSJ1352+3216 &   BELLS & 0.463 & 1.034 & $0.58\pm0.06$ & 11.37 & 0.86 \\
  SDSSJ1522+2910 &   BELLS & 0.555 & 1.311 & $0.89\pm0.09$ & 11.39 & 0.74 \\
  SDSSJ1541+1812 &   BELLS & 0.560 & 1.113 & $0.76\pm0.08$ & 11.41 & 0.93 \\
  SDSSJ1542+1629 &   BELLS & 0.352 & 1.023 & $0.73\pm0.07$ & 11.52 & 0.81 \\
  SDSSJ1545+2748 &   BELLS & 0.522 & 1.289 & $2.59\pm0.26$ & 11.82 & 0.42 \\
  SDSSJ1601+2138 &   BELLS & 0.543 & 1.446 & $0.44\pm0.04$ & 11.40 & 0.91 \\
  SDSSJ1611+1705 &   BELLS & 0.477 & 1.211 & $1.00\pm0.10$ & 11.23 & 0.74 \\
  SDSSJ1631+1854 &   BELLS & 0.408 & 1.086 & $1.43\pm0.14$ & 11.94 & 0.88 \\
  SDSSJ1637+1439 &   BELLS & 0.391 & 0.874 & $1.04\pm0.10$ & 11.22 & 0.75 \\
  SDSSJ2122+0409 &   BELLS & 0.626 & 1.452 & $0.90\pm0.09$ & 11.33 & 0.63 \\
  SDSSJ2125+0411 &   BELLS & 0.363 & 0.978 & $0.90\pm0.09$ & 11.68 & 0.82 \\
  SDSSJ2303+0037 &   BELLS & 0.458 & 0.936 & $1.35\pm0.14$ & 11.62 & 0.39 \\
   HE0047$-$1756 & CASTLES & 0.408 & 1.670 & $0.49\pm0.12$ & 11.18 & 0.80 \\
     Q0142$-$100 & CASTLES & 0.491 & 2.719 & $0.51\pm0.03$ & 11.53 & 1.18 \\
   QJ0158$-$4325 & CASTLES & 0.317 & 1.294 & $0.66\pm0.08$ & 11.07 & 0.58 \\
   HE0230$-$2130 & CASTLES & 0.522 & 2.162 & $0.14\pm0.15$ & 10.82 & 0.87 \\
SDSSJ0246$-$0825 & CASTLES & 0.723 & 1.686 & $0.18\pm0.06$ & 11.08 & 0.53 \\
     MG0414+0534 & CASTLES & 0.958 & 2.639 & $0.78\pm0.02$ & 11.72 & 1.11 \\
   HE0435$-$1223 & CASTLES & 0.454 & 1.689 & $0.76\pm0.04$ & 11.38 & 1.22 \\
       B0712+472 & CASTLES & 0.406 & 1.339 & $0.36\pm0.03$ & 11.13 & 0.72 \\
     MG0751+2716 & CASTLES & 0.350 & 3.200 & $0.31\pm0.04$ & 10.22 & 0.42 \\
     HS0818+1227 & CASTLES & 0.390 & 3.115 & $0.62\pm0.05$ & 11.22 & 1.37 \\
       B0850+054 & CASTLES & 0.588 & 3.930 & $0.16\pm0.01$ & 10.64 & 0.34 \\
  SDSSJ0924+0219 & CASTLES & 0.393 & 1.523 & $0.30\pm0.02$ & 11.06 & 0.88 \\
 LBQS1009$-$0252 & CASTLES & 0.871 & 2.739 & $0.19\pm0.04$ & 10.89 & 0.77 \\
      J1004+1229 & CASTLES & 0.950 & 2.640 & $0.34\pm0.24$ & 11.17 & 0.83 \\
       B1030+074 & CASTLES & 0.599 & 1.535 & $0.23\pm0.06$ & 11.07 & 0.91 \\
   HE1104$-$1805 & CASTLES & 0.729 & 2.303 & $0.64\pm0.20$ & 11.44 & 1.40 \\
      PG1115+080 & CASTLES & 0.311 & 1.736 & $0.46\pm0.03$ & 11.10 & 1.14 \\
  RXJ1131$-$1231 & CASTLES & 0.295 & 0.658 & $1.13\pm0.21$ & 11.34 & 1.83 \\
  SDSSJ1138+0314 & CASTLES & 0.445 & 2.442 & $0.19\pm0.03$ & 10.92 & 0.57 \\
  SDSSJ1155+6346 & CASTLES & 0.176 & 2.888 & $0.43\pm0.05$ & 10.86 & 0.76 \\
SDSSJ1226$-$0006 & CASTLES & 0.517 & 1.126 & $0.45\pm0.07$ & 11.21 & 0.57 \\
   LBQS1333+0113 & CASTLES & 0.440 & 1.571 & $0.31\pm0.02$ & 11.07 & 0.85 \\
    Q1355$-$2257 & CASTLES & 0.702 & 1.373 & $1.24\pm0.29$ & 11.56 & 0.62 \\
   HST14113+5211 & CASTLES & 0.465 & 2.811 & $0.47\pm0.04$ & 10.90 & 0.84 \\
   HST14176+5226 & CASTLES & 0.809 & 3.400 & $0.70\pm0.04$ & 11.25 & 1.41 \\
       B1422+231 & CASTLES & 0.337 & 3.620 & $0.32\pm0.03$ & 10.83 & 0.78 \\
     MG1549+3047 & CASTLES & 0.111 & 1.170 & $0.82\pm0.01$ & 10.98 & 1.15 \\
       B1608+656 & CASTLES & 0.630 & 1.394 & $0.64\pm0.05$ & 11.67 & 0.81 \\
 PMNJ1632$-$0033 & CASTLES & 1.165 & 3.424 & $0.20\pm0.03$ & 10.71 & 0.64 \\
    FBQ1633+3134 & CASTLES & 0.684 & 1.518 & $2.93\pm1.26$ & 12.07 & 0.35 \\
     MG1654+1346 & CASTLES & 0.254 & 1.740 & $0.89\pm0.02$ & 11.29 & 1.05 \\
       B1938+666 & CASTLES & 0.881 & 2.059 & $0.69\pm0.05$ & 11.17 & 0.50 \\
      MG2016+112 & CASTLES & 1.004 & 3.273 & $0.22\pm0.02$ & 11.41 & 1.78 \\
  WFI2033$-$4723 & CASTLES & 0.661 & 1.660 & $0.72\pm0.09$ & 11.46 & 1.12 \\
       B2045+265 & CASTLES & 0.867 & 1.280 & $0.41\pm0.04$ & 11.16 & 1.06 \\
   HE2149$-$2745 & CASTLES & 0.603 & 2.033 & $0.50\pm0.09$ & 11.22 & 0.86 \\
       Q2237+030 & CASTLES & 0.039 & 1.695 & $3.86\pm0.09$ & 11.08 & 0.90 \\
 COSMOS5921+0638 & CASTLES & 0.551 & 3.140 & $0.41\pm0.01$ & 11.12 & 0.72 \\
  SDSSJ0743+2457 & SQLS+AO & 0.381 & 2.165 & $0.09\pm0.04$ & 10.77 & 0.56 \\
  SDSSJ0806+2006 & SQLS+AO & 0.573 & 1.538 & $0.23\pm0.01$ & 11.17 & 0.76 \\
  SDSSJ0819+5356 & SQLS+AO & 0.294 & 2.239 & $1.12\pm0.06$ & 11.89 & 2.06 \\
  SDSSJ0946+1835 & SQLS+AO & 0.388 & 4.799 & $0.75\pm0.04$ & 11.63 & 1.45 \\
  SDSSJ1055+4628 & SQLS+AO & 0.388 & 1.249 & $0.28\pm0.03$ & 10.95 & 0.58 \\
  SDSSJ1313+5151 & SQLS+AO & 0.194 & 1.877 & $0.57\pm0.03$ & 10.94 & 0.60 \\
  SDSSJ1620+1203 & SQLS+AO & 0.398 & 1.158 & $0.90\pm0.05$ & 11.47 & 1.40 \\
SL2SJ0213$-$0743 &    SL2S & 0.717 & 3.480 & $2.45\pm0.25$ & 11.96 & 2.39 \\
SL2SJ0214$-$0405 &    SL2S & 0.609 & 1.880 & $0.93\pm0.09$ & 11.72 & 1.41 \\
SL2SJ0217$-$0513 &    SL2S & 0.646 & 1.850 & $0.61\pm0.06$ & 11.69 & 1.27 \\
SL2SJ0219$-$0829 &    SL2S & 0.389 & 2.150 & $0.57\pm0.06$ & 11.54 & 1.30 \\
SL2SJ0225$-$0454 &    SL2S & 0.238 & 1.200 & $2.28\pm0.23$ & 11.76 & 1.76 \\
SL2SJ0226$-$0420 &    SL2S & 0.494 & 1.230 & $1.06\pm0.11$ & 11.71 & 1.19 \\
SL2SJ0232$-$0408 &    SL2S & 0.352 & 2.340 & $0.96\pm0.10$ & 11.34 & 1.04 \\
SL2SJ0849$-$0412 &    SL2S & 0.722 & 1.540 & $0.49\pm0.05$ & 11.69 & 1.10 \\
SL2SJ0849$-$0251 &    SL2S & 0.274 & 2.090 & $1.46\pm0.15$ & 11.44 & 1.16 \\
SL2SJ0901$-$0259 &    SL2S & 0.670 & 1.190 & $0.50\pm0.05$ & 10.97 & 1.03 \\
SL2SJ0904$-$0059 &    SL2S & 0.611 & 2.360 & $2.50\pm0.25$ & 11.58 & 1.40 \\
  SL2SJ0959+0206 &    SL2S & 0.552 & 3.350 & $0.54\pm0.05$ & 11.22 & 0.74 \\
  SL2SJ1359+5535 &    SL2S & 0.783 & 2.770 & $1.76\pm0.18$ & 11.35 & 1.14 \\
  SL2SJ1405+5243 &    SL2S & 0.526 & 3.010 & $0.73\pm0.07$ & 11.72 & 1.51 \\
  SL2SJ1406+5226 &    SL2S & 0.716 & 1.470 & $0.60\pm0.06$ & 11.55 & 0.94 \\
  SL2SJ1411+5651 &    SL2S & 0.322 & 1.420 & $0.65\pm0.07$ & 11.38 & 0.93 \\
  SL2SJ1420+5258 &    SL2S & 0.380 & 0.990 & $1.04\pm0.10$ & 11.47 & 0.96 \\
  SL2SJ1420+5630 &    SL2S & 0.483 & 3.120 & $1.31\pm0.13$ & 11.82 & 1.40 \\
  SL2SJ1427+5516 &    SL2S & 0.511 & 2.580 & $0.50\pm0.05$ & 11.27 & 0.81 \\
  SL2SJ2203+0205 &    SL2S & 0.400 & 2.150 & $0.72\pm0.07$ & 11.36 & 1.95 \\
SL2SJ2213$-$0009 &    SL2S & 0.338 & 3.450 & $0.50\pm0.05$ & 10.91 & 1.07 \\
 \hline
\end{longtable}

\label{lastpage}

\end{document}